# Specific features of electron spin resonance in quasi-1D magnet $\beta$-$TeVO_4$

Yu. Savina[1,*], A. Bludov[1], V. Pashchenko[1], S. Gnatchenko[1], A. Stepanov[2], P. Lemmens[3]
[1] B.I Verkin Institute for Low Temperature Physics and Engineering NASU, Lenin ave., 47, Kharkiv 61103, Ukraine
[2] Insitut Matériaux Microélectronique Nanosciences de Provence, CNRS UMR 6242, FST, Aix-Marseille Université, Marseille Cedex 20 F-13397, France
[3] Institute for Condensed Matter Physics, TU Braunschweig, Braunschweig D-38106, Germany
*E-mail: savina@ilt.kharkov.ua

The angular and temperature dependences of single crystal ESR spectra at X-band frequency of quasi-1D spin ½ antiferromagnetic zigzag chain system $\beta$-$TeVO_4$ are reported. Two resonance components in the ESR spectrum have been detected: a weak narrow resonance on the top of the extremely broad signal. The narrow line might be due to an impurity, the characteristics of the broad absorption do not correlate with a picture of the typical behavior for the $V^{4+}$ ions ($S = ½$) with a square-pyramidal environment. It was found that at low temperatures the divergence of the spin susceptibility and linewidth of the observed excitations can be due to a realization of three-dimensional magnetic ordering.

**PACS**: 76.30.-v, 76.30.Fc, 75.50.Ee, 75.10.Pq, 75.30.Gw

**Keywords**: electron spin resonance, antiferromagnets, spin chain systems, magnetic anisotropy, *g*-factors

## 1. Introduction

The properties of various one-dimensional (1D) quantum spin systems have been studied intensively both experimentally and theoretically [1-7]. The physical realizations of 1D spin systems are compounds in which the exchange interaction $J$ along one direction is significantly stronger than the exchange $J'$ along to other direction ($J >> J'$ or $J'/J << 1$). Although in 1D systems a classical Neel ordered state isn't expected, but the existence of a small non-vanishing interchain coupling often lead to the realization of a 3D long-range magnetically ordered state at low temperatures.

The vanadium oxides containing $VO_5$ square pyramids and forming a large variety of the different chain-like structures are of particular importance since they are expected to be new spin systems with pronounced 1D or quasi-1D magnetic properties. For example such quantum spin systems are: uniform linear chain ($\alpha'$-$NaV_2O_5$ [1]) and alternating AF or FM exchange chain ($(VO)_2P_2O_7$ [2], $\alpha$-$TeVO_4$ [3]), zigzag chain ($LiV_2O_5$ [4], $CdVO_3$ [5]), two-leg spin ladder ($CaV_2O_5$ [6], $\beta$-$SrV_6O_{15}$ [7]) etc.

Recently a quasi-1D magnetic behavior was clearly revealed in the zigzag chain compound $\beta$-$TeVO_4$ by means magnetic susceptibility $\chi(T)$ measurements in single crystal samples [8]. It was shown the magnetic data might be well ascribed by using a spin $S = ½$ antiferromagnetic (AFM) Heisenberg chain model with only one dominant AFM nearest-neighbor exchange coupling $J / k_B = 21.4$ K and much weaker other couplings (intra- and interchain, less than 2 K) that lead to three-dimentional AFM ordering at the Néel temperature $T_N = 4.65$ K. In addition a cascade of several phase transformations was found at $T < T_N$ and claimed that this system perhaps may have a helimagnetic long-range ordering at low temperatures. Magnetic data shows an axial anisotropy of magnetic properties with respect to the crystallographic *b*-axis, that may be due to an axial symmetry of the *g*-tensor for the $V^{4+}$ ions with a square-pyramidal oxygen environment. High-temperature estimates of *g*-values are in a satisfactory agreement with the expected ones (1.94−1.98) from the literature [9]. At the same time the obtained $g$-tensor according to uniform AFM chain model [8] give much bigger values ($g_{\perp} = 2.19, 2.20$: $g_{\|} = 2.28$) that requires a clarification by accurate spectroscopic methods of investigation.

The present paper is the first report on results of electron spin resonance (ESR) studies on single crystals of $\beta$-$TeVO_4$ by using a standard X-band spectrometer in a wide temperature range. The aim of this work is to study the particularities of magnetic resonance behavior and to obtain information about the anisotropy of the $g$-tensor and the linewidth broadening for the $V^{4+}$ ions ($S = ½$) in zigzag chain system $\beta$-$TeVO_4$.

## 2. Experimental details

The growth of $\beta$-$TeVO_4$ single crystals was described elsewhere [10]. Single crystal sample with dimensions of $1{\times}1{\times}4$ $mm^3$ ($m = 20.45$ mg) was oriented by using a X-ray Laue diffractometer. Accuracy of crystal orientation was better than ±1°.

The ESR measurements (the temperature and angular dependences) on single crystal $\beta$-$TeVO_4$ were performed by using a EMX Bruker spectrometer with a standard $TE_{102}$ cavity operating at X-band frequency 9.397 GHz in the temperature range $4.7\ K \leq T \leq 300\ K$ and magnetic fields from 0 to $1.2{\cdot}10^4$ Oe. To improve the signal-to-noise ratio, the derivatives of absorption $dI / dH$ was detected by the lock-in technique with field modulation frequency of 100 kHz. Derivatives of the resonance absorption lines were recorded. A single crystal sample was mounted on a quartz stick (sample holder) and the stick was rotated around the axis perpendicular to the applied static magnetic field. The angular dependence of ESR spectra was recorded in the range [−90º; 90º] with an angular step width of 20º from orientation $H \perp b$ to $H \| b$ and again to $H \perp b$. The angle between $b$-axis and the direction of the static magnetic field $H$ was controlled by a goniometer with a precision better than ±0.25º. The temperature was controlled between 4.7 K and room temperature by a He-gas continuous-flow cryostat (Oxford Instruments) with the temperature stability of ±0.1 K. The etalon sample of DPPH with an effective $g$-value of 2.0032 was used as a field marker.

Each recorded spectrum always was corrected by subtraction of the background signal measured at the same experimental conditions (without sample) as needed for detection of very weak and/or broad signals.

## 3. Results

### 3.1 Structure and important parameters

As described in Ref. [10] the compound $\beta$-$TeVO_4$ crystallizes in the monoclinic system with the space group $P2_1/c$ and the parameters: $a = 4.379 \pm 0.002$ Å, $b = 13.502 \pm 0.004$ Å, $c = 5.446 \pm 0.002$ Å and $\beta = 91.72° \pm 0.05°$ with Z = 4 (f.u. $TeVO_4$ per unit cell). The structure is composed of zigzag vanadium chains parallel to the crystallographic $c$-axis. A view of zigzag chain in $\beta$-$TeVO_4$ is shown in Fig. 1. Along the chain direction, the slightly distorted square pyramids $VO_5$ share corners and induce the nearest-neighbor coupling $J_{NN}$ mediated by V–O–V superexchange interaction. As it was demonstrated for another vanadium compounds [5, 6] the corner-sharing exchange coupling $J_{NN}$ between vanadium ions is expected to be weak in magnitude. The possible next-nearest-neighbor exchange path V–O–O–V through two oxygens of the basal plane of the $VO_5$ pyramid is forming a zigzag chain as two-leg ladder and leading to the competition (perhaps frustration) between the nearest-neighbor coupling $J_{NN}$ and the next-nearest-neighbor exchange $J_{NNN}$. Thus, an effective 1D magnetic model for $\beta$-$TeVO_4$ can be regarded as the isolated two-leg ladder (double chain) with two different exchange integrals ($J_1$-$J_2$ model). Note, there are two identical zigzag chains, in which all apices of square pyramids are pointing alternatively below and above the (*bc*)-plane. The $Te^{4+}$ cations (are not shown in Fig. 1) are positioned between the chains and the magnetic interactions between the chains are believed to be very weak. A twofold axis $C_2$ is directed along the *b*-axis. The nearest intrachain distance is V−V = 3.6427 Å. The nearest V−V distances perpendicular to the chain direction are 4.9149 (along the *b*-axis) and 4.3790 Å (along the *a*-axis), respectively.

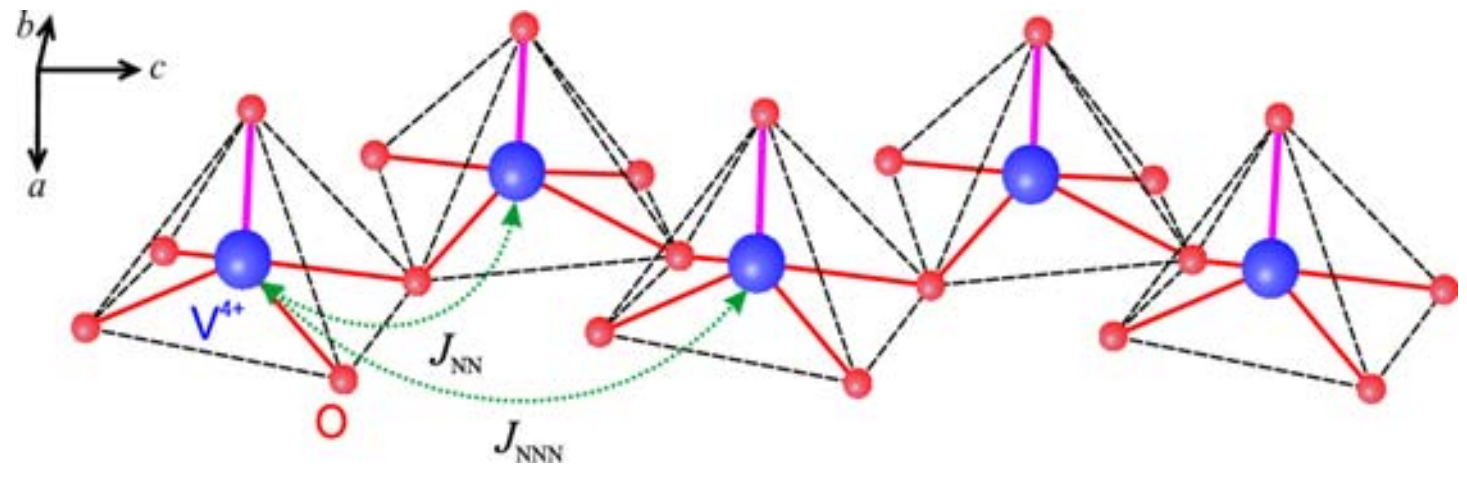

**Figure 1**. A view of zigzag chain of $\beta$-$TeVO_4$ formed by slightly distorted square pyramids $VO_5$ sharing corners.

It is well known that a 1D Heisenberg antiferromagnet shows a characteristic maximum at $T_{max}$ in the magnetic susceptibility and the ESR intensity as well, which suggests the development of a short-range correlations at low temperatures [11]. Due to this development of the short-range spin-spin correlations, electron spin resonance of 1D Heisenberg antiferromagnet typically shows two characteristic features. One is the *g*-shift with decreasing temperature, another feature is the linewidth broadening [12]. Both phenomena will be in a focus of our interest for the current investigation.

## *3.2. Temperature dependence*

At different temperatures the ESR spectra were recorded on a single crystal $\beta$-$TeVO_4$ in order to study the variation in intensities and the linewidths of the detected absorption signals with the temperature.

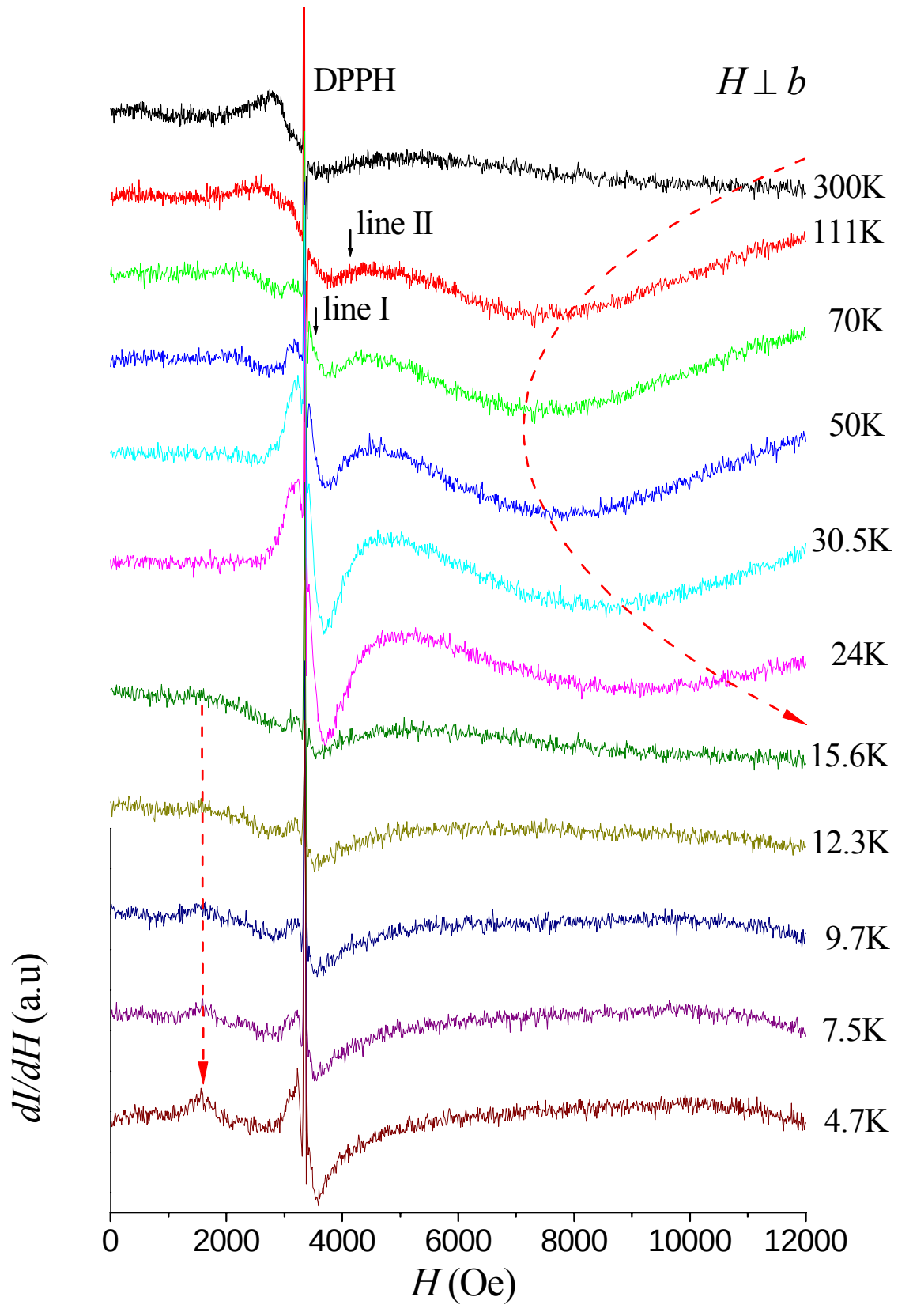

**Figure 2**. The temperature dependence of the ESR spectra of single crystal $\beta$-$TeVO_4$ for $H \perp b$.

Orientation $H \perp b$ was chosen by us because it was observed the largest quantity of features in the spectra. It should be noted that at the beginning of our experiment a few times we had to increase the mass of the sample (from ~1 mg up to ~20 mg) since the observed absorption intensity at room temperature was too low, that is not typical for vanadium compounds with the $V^{4+}$ ions. Figure 2 shows the temperature evolution of the ESR spectra of single crystal $\beta$-$TeVO_4$ at 9.397 GHz from room temperature down to 4.7 K and the orientation of applied magnetic field perpendicular to the *b*-axis. One can distinguish two qualitatively different behaviors. Each ESR spectrum shows the presence of a small resonance absorption line (denoted as line I) around a resonance field $H_{res} \approx 3445$ Oe corresponding to $\langle g \rangle \approx 1.958$. This signal can be observed for all orientations and temperatures. The halfwidth at half-maximum of this small resonance is about $\Delta H \sim 1.5$ kOe at room temperature. As one see in Fig. 2, with temperature decreasing the intensity of line I is monotonously increasing down to ~20 K, than there is a rapid drop, but the weak signal remains and grows at further lowering temperature. For $H \perp b$, the second feature as an additional very broad resonance ($\Delta H \sim 20$ kOe, $T = 300$ K, denoted as line II) appears, which is centered around $H_{res} \approx 4$ kOe. The linewidth of line II shows the significant temperature dependence from 300 K down to ~ 15 K. The temperature evolution of the narrowing and broadening process for line II is shown schematically as a dash line in Fig. 2. It should be noted, the broad line II reaches its minimal linewidth value around $\Delta H \approx 10$ kOe at ~70 K. In contrast to line I this very broad resonance becomes unobservable below ~15 K. Additionally, below 15 K the small signal at half-field position $H = ½H_{res}$ is detected (marked as a vertical dash line). The narrowest line at $H = 3366.6$ Oe in ESR spectra is the etalon signal of DPPH.

### 3.3. Angular dependence

In order to study the anisotropy of two observed lines we have measured the angular dependence of the ESR spectra of $\beta$-$TeVO_4$ at $T = 31.4$ K, at which the both absorption intensities are suitable for monitoring. Orientation of external magnetic field changed in one plane from $H \perp b$ to $H \parallel b$ and again to $H \perp b$ ([−90º; 90º]) with an angular step of 20º. The angular transformation of the ESR spectra of single crystal $\beta$-$TeVO_4$ are shown in Fig. 3. From each spectrum the etalon signal of DPPH was subtracted for clarity. As one can see in Fig. 3 the broad line II is almost non-detectable for $H \parallel b$ (its linewidth becomes extremely large). At the same time the linewidth of the line I is weak angular dependent. The signal for line I, which is observed for all orientations with a nearly invariable linewidth and intensity, satisfactorily agrees with the expected very weak anisotropy for the ions with a $S$-state. In our experiments it was confirmed that the resonance properties of single crystal $\beta$-$TeVO_4$ have an axial symmetry with the symmetry axis coinciding with the direction of the $b$-axis as was claimed from magnetic measurements [8].

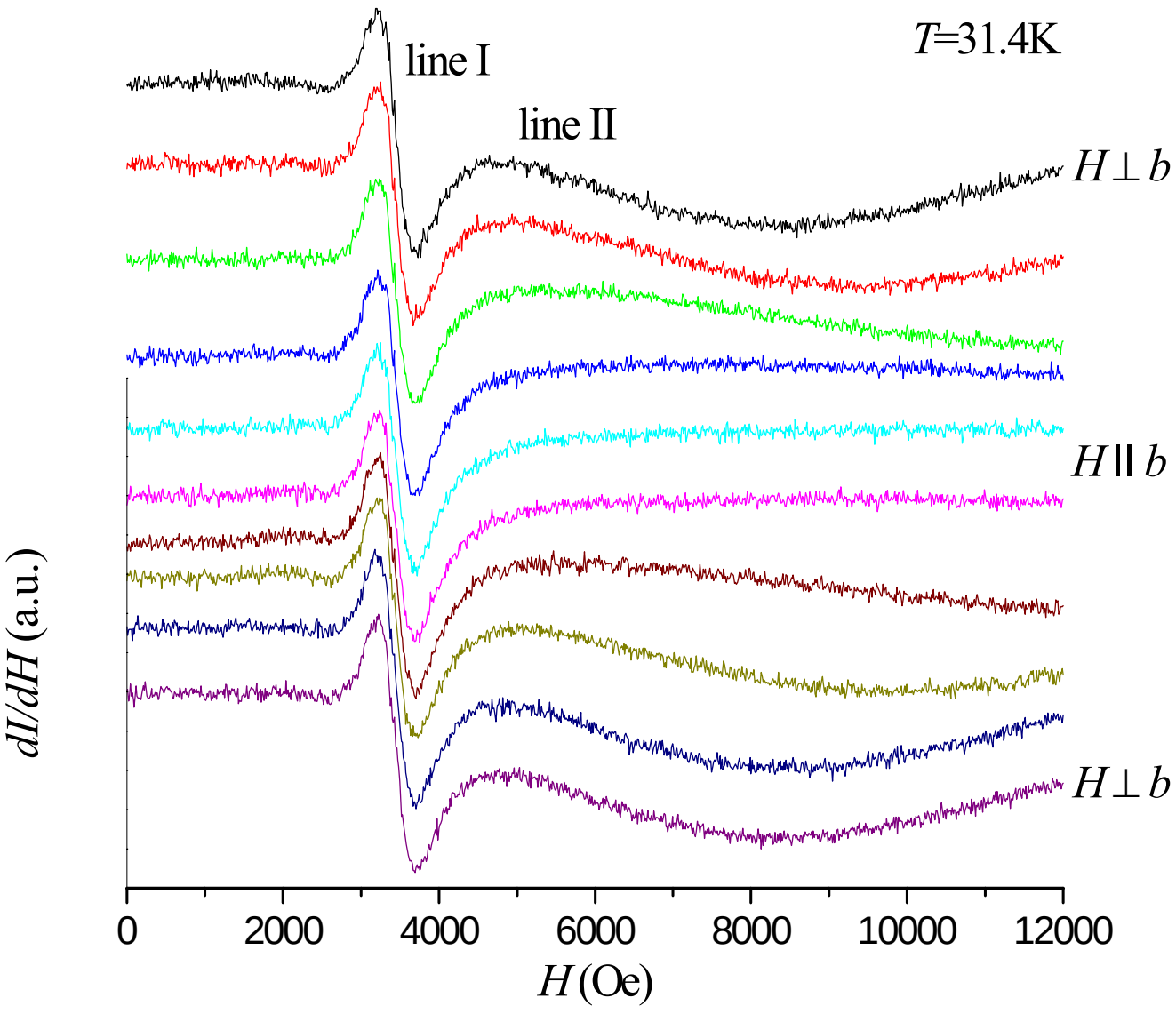

**Figure 3.** The angular dependence of the ESR spectra of single crystal $\beta$-$TeVO_4$ at $T = 31.4$ K.

## 4. Discussion

For obtaining the effective parameters of the resonance absorptions such as a spin susceptibility $\chi_{ESR}$, linewidth $\Delta H$ and resonance position $H_{res}$ we have used a simple model containing of two different lines with a Lorentzian line shape:

$$\frac{dI}{dH} = \frac{d}{dH}\left[\frac{2/\pi\, \chi^{I}_{ESR}\Delta H_{I}}{4(H-H^{I}_{res})^2+\Delta H_{I}^{2}} + \frac{2/\pi\, \chi^{II}_{ESR}\Delta H_{II}}{4(H-H^{II}_{res})^2+\Delta H_{II}^{2}}\right],$$

where $H^{i}_{res}$ is a center of the resonance peak, $\Delta H_i$ is a full linewidth between half-amplitude points of the nonderived ESR line and $\chi^{i}_{ESR}$ is a spin susceptibility (an area under the curve from the baseline) for two observed lines $i$ = I, II.

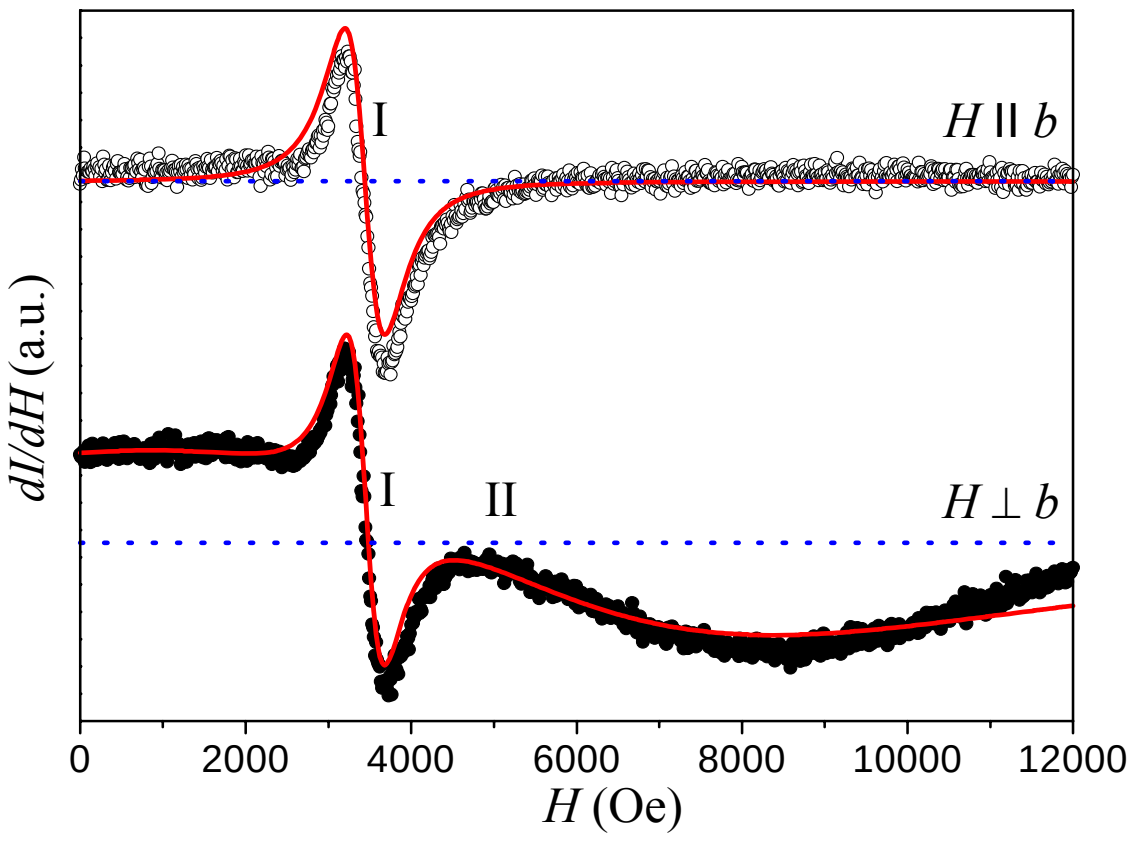

**Figure 4**. Two ESR spectra of single crystal $\beta$-$TeVO_4$ for $H \parallel b$ (○) and $H \perp b$ (●) at $T = 31.4$ K. The solid lines were calculated by using a model with two Lorentzian shape lines.

As an example, two ESR spectra obtained at $T = 31.4$ K for $H \parallel b$ (open circles) and $H \perp b$ (solid circles) together with the corresponding best fits (solid lines) by using the model mentioned above are shown in Fig. 4. The dot lines show a zero intensity level (where $dI / dH = 0$). The best fit parameters are summarized in Table 1. As one can see in Fig. 4, all spectra is satisfactorily described by previously proposed equation, adding an extra Lorentzian line of the small resonance on the top of the very broad line. The successful fitting of any spectrum obtained at X-band frequency justify an approximate treatment of the ESR data by means of this model for a restoration of the temperature and angular dependences for spectroscopic characteristics.

| | $H_{res}$, (Oe) | | $\Delta H$, (Oe) | | $\chi_{ESR}$, (a.u.) | |
|---|---|---|---|---|---|---|
| | $H \parallel b$ | $H \perp b$ | $H \parallel b$ | $H \perp b$ | $H \parallel b$ | $H \perp b$ |
| line I | 3442 | 3450 | 860 | 771 | $2.26 \cdot 10^{9}$ | $2.07 \cdot 10^{9}$ |
| line II | 8767 | 4541 | ~85000 | 13350 | $\sim 5.21 \cdot 10^{11}$ | $3.11 \cdot 10^{11}$ |

**Table 1.** The best fit parameters for a model with two Lorentzian shape lines. The ESR spectra of $\beta$-$TeVO_4$ were taken for orientations $H \parallel b$ and $H \perp b$ at $T = 31.4$ K.

The obtained results of analysis for angular and temperature dependences of single crystal $\beta$-$TeVO_4$ are summarized in Fig. 5 and Fig. 6. Error bars on some points indicate the uncertainty in a reported measurement.

In Fig. 5, we show the angular variation of the field position $H_{res}(\theta)$ and the linewidth $\Delta H(\theta)$ for two absorption lines in $\beta$-$TeVO_4$ at $T = 31.4$ K in a rotation of the magnetic field $H$ about the crystallographic $b$-axis, where $\theta$ is the angle between the static magnetic field and the $b$-axis. As one can see in Fig. 5 both lines exhibit a 180° symmetry as expected. It should be noted for $H \parallel b$ the resonance field $H_{res}$ of the line I is minimal, when the resonance field $H_{res}$ of the line II has a maximum. The angular variation of $H_{res}(\theta)$ for line I is rather weak: resonance field changes from 3440 Oe to 3450 Oe. The field position $H_{res}(\theta)$ for line II is larger than for line I: it changes from 4500 Oe up to 9000 Oe. The angular behavior of linewidths are similar for both excitations except the difference in a magnitude. The anisotropy of the $g$-tensor in the ($ac$)-plane perpendicular to the $b$-axis is much smaller. It can be limited by experimental accuracy to a value smaller then 0.002. Overall the experimental data reveal the typical pattern due to an axial anisotropy of the resonance field, which allows to determine the $g$-values for $g_{\parallel}$ and $g_{\perp}$ and the linewidths $\Delta H_{\parallel}$ and $\Delta H_{\perp}$ for magnetic field applied parallel or perpendicular to the local symmetry axis, respectively.

Using the resonance condition at the frequency $v$ the effective $g$-factor can be determined from the resonance field $H_{res}(\theta)$ as

$$g(\theta) = hv / (\mu_B H_{res}(\theta)),$$

where $h$ is the Planck constant and $\mu_B$ is the Bohr magneton. Fitting the angular dependencies to the equation

$$g(\theta) = \sqrt{g_{\parallel}^2 \cos^2\theta + g_{\perp}^2 \sin^2\theta}$$

yields parallel and perpendicular $g$-factors with respect to the $b$-axis: $g_{\perp} = 1.955 \pm 0.002$ and $g_{\parallel} = 1.960 \pm 0.002$ for line I; $g_{\perp} = 1.60 \pm 0.05$ and $g_{\parallel} = 0.80 \pm 0.05$ for line II. The first pair (for line I) is found to be close to the spin-only value $g = 2$, which is typical for 3d ions, where the orbital momentum is quenched. The second one with the average value $\langle g \rangle \approx 1.2$ (line II) is not typical for one particle approximation for $V^{4+}$ magnetic centers.

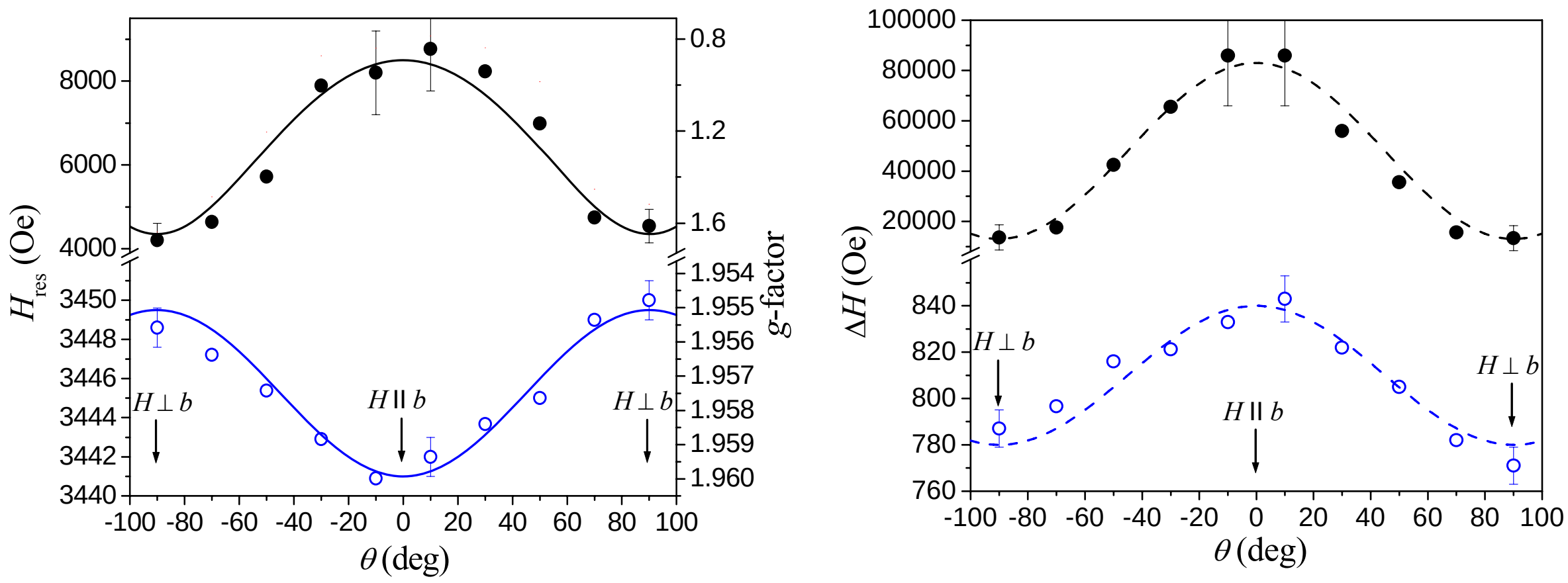

**Figure 5.** The angular dependencies of the position $H_{res}(\theta)$ (left) and linewidth $\Delta H(\theta)$ (right) of two absorption lines of $\beta$-$TeVO_4$ at $T = 31.4$ K. Data for line I is shown as open circles; for the line II – solid circles.

In the next step, the angular dependence of the ESR linewidth was analyzed (see the right panel in Fig. 5). At given temperature, the angular dependence of the linewidth seems to be described by the following empirical equation

$$\Delta H(\theta) = \Delta H_{\perp} \sin^2\theta + \Delta H_{\parallel} \cos^2\theta$$

by using two fitting parameters $\Delta H_{\perp}$ and $\Delta H_{\parallel}$. In the right panel in Fig. 5 the dash lines represent the best fit results according to this equation with the parameters: $\Delta H_{\perp}^{I} = 780\pm10$ Oe and $\Delta H_{\parallel}^{I} = 840\pm10$ Oe and $\Delta H_{\perp}^{II} = 13\pm2$ kOe and $\Delta H_{\parallel}^{II} = 83\pm20$ kOe for the line I and II, respectively. This anisotropy is qualitatively different from expected one for one-dimensional spin systems [13], where $\Delta H(\theta) \propto (3\cos^2\theta - 1)$.

The obtained temperature dependences of spin susceptibility $\chi_{ESR}(T)$, linewidth $\Delta H(T)$ and resonance field $H_{res}(T)$ of two observed lines in single crystal $\beta$-$TeVO_4$ for measuring geometry $H \perp b$ are shown in Fig. 6.

Spin susceptibility $\chi_{ESR}(T)$ of both lines are diverged at low temperatures (see the upper panels in Fig. 6). It seems that the ESR intensity follows a Curie-Weiss law

$$\chi_{ESR}(T) = C/(T - \Theta).$$

We have estimated a Curie-Weiss temperature as $\Theta \sim 5\div9$ K, that is rather closed to the temperature of magnetic ordering $T_N = 4.65$ K. It should be noted the temperature dependence of the integral intensity for both lines is satisfactorily agreement with the results of the magnetic susceptibility measurements for $\beta$-$TeVO_4$ [8]. Since it was a limited number of the ESR experiments at low temperatures and the accuracy of spectroscopic information should be higher, the corresponding characteristic susceptibility maximum of 1D magnetic system as claimed by magnetic measurements (at 14 K for $\beta$-$TeVO_4$) was not reliably observed.

The obtained intensity ratio $\chi_{ESR}^{I} / \chi_{ESR}^{II}$ between the weak narrow and very broad resonance is approximately $10^{-3}$. This demonstrates that the broad absorption line represents by far the major part of the total ESR intensity corresponding to the spin susceptibility of the compound at least above 15 K. This dominating signal is strongly suppressed due to the linewidth broadening for $H \parallel b$. The exact origin of the main component of the ESR spectrum as a very broad resonance and its large field shift is not clear yet. The extremely large linewidth could be due to the very poor quality of the single crystal $\beta$-$TeVO_4$, but it is completely undetectable and is not supported by any structural or magnetic investigations. We are sure that we have examined sufficiently high quality single crystals. Possible physical reasons for the anomalously large broadening could be a frustration phenomenon in one-dimensional spin system with complex topology of interactions or the presence of the antisymmetric exchange interactions of Dzyaloshinskii-Moriya type, which are very often used recently to explain the nature of the very broad resonance lines. At the current time we do not have enough grounds to make a definite conclusion, and we would only emphasize that the resonant properties of $\beta$-$TeVO_4$ is an exceptional example of a compound with the $V^{4+}$ ions, which is quite different from other known vanadium containing systems. Perhaps it will be very useful to study the resonance properties of this compound at low-temperature, which is planned in the nearest future. But even the detailed temperature dependence of $\chi_{ESR}$ and $\Delta H$ has not been studied for $\beta$-$TeVO_4$ near the Neel temperature in the present experiments. We do not exclude a possibility that the existence of a weak resonance (line I) may be due to the presence of small amounts of impurities or defects in a crystal (at the level of 0.1%), but this question requires further detailed studies. Sufficiently convincing arguments in favor of what we can deal with the impurity signal may be, for example, the following two facts: i) its intensity is very weak compared to the second (main) component of the spectrum, and ii) the obtained anisotropy (1.955−1.96) is noticeably smaller than the expected one (1.94−1.98) for the $V^{4+}$ ions with a square pyramidal environment. Our preliminary attempts to record the ESR spectra at higher frequencies (above and below $T_N$) demonstrate a complexity of the resonance situation for given compound and the resonance spectra always contain the multiple excitations with the $g$-factors close to 2 and/or to 1.2.

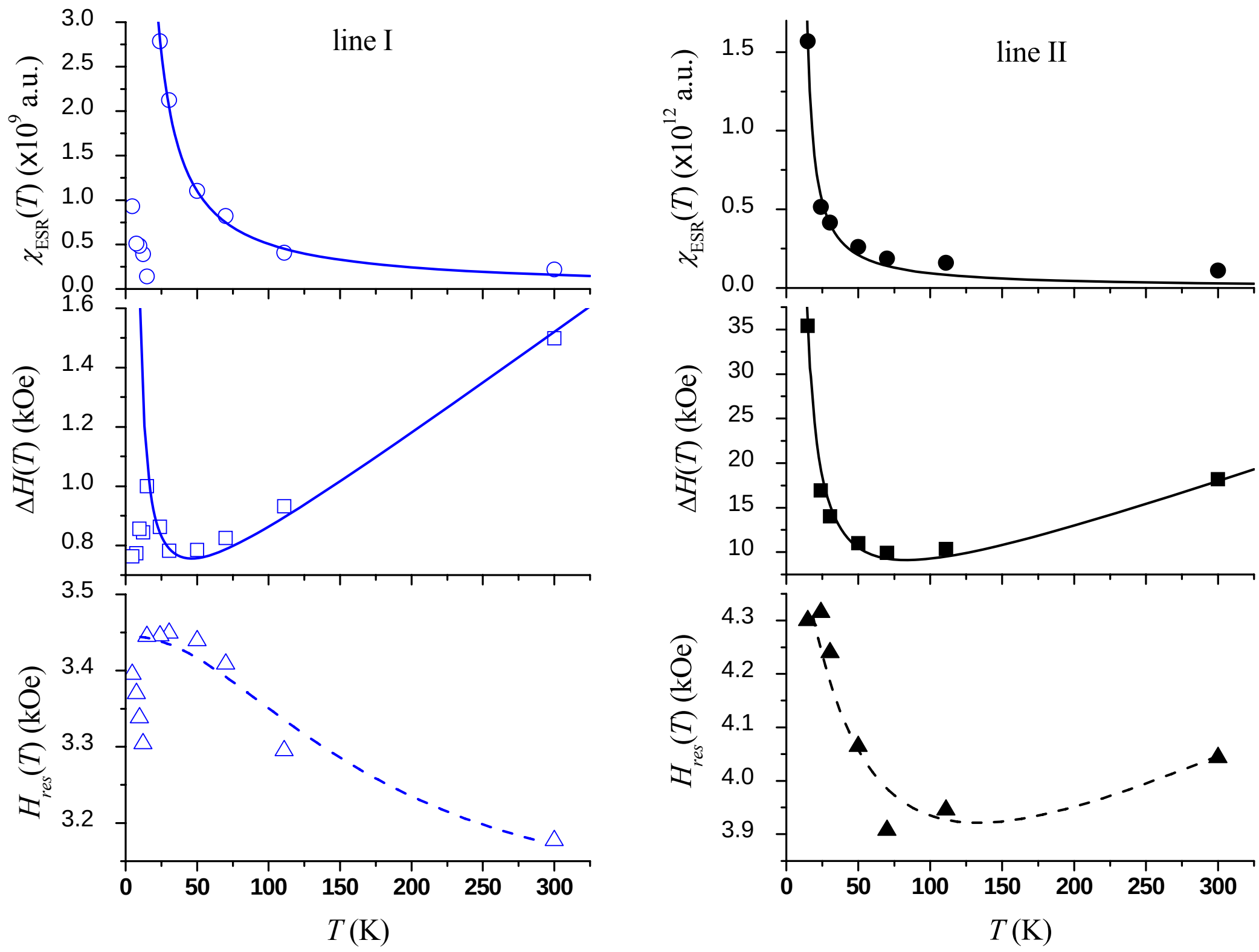

**Figure 6**. Temperature dependences of the spin susceptibility $\chi_{ESR}(T)$ (circle), linewidth $\Delta H(T)$ (square) and resonance field $H_{res}(T)$ (triangle) of two ESR lines in $\beta$-$TeVO_4$ for $H \perp b$. Open and solid symbols correspond to the data for line I and line II, respectively.

The middle panels in Fig. 6 clearly demonstrate that with temperature decreasing from 300 K down to ~ 50 K (~ 75 K for line II) the linewidth $\Delta H$ of the both lines reduces, at ~ 50 K (~ 75 K) the linewidth $\Delta H$ reaches a minimum and then it undergoes a rapid broadening again. A drastic ESR line broadening at low temperatures can be attributed to the rapid development of the intrachain spin correlations with a subsequent growth of critical 3D correlations due to the vicinity of Neel's temperature. In low-dimensional spin systems an essential broadening of the ESR line can occur in a rather broad temperature range approximately up to $10T_N$. Taking into account $T_N = 4.65$ K the development of the critical behavior of $\Delta H$ might be begin at ~ 50 K, that is comparable with the observations for our system. When the temperature is lowered towards the critical temperature $T_{cr} \equiv T_N$ a power low $\Delta H_{cr} \propto \left(T - T_{cr}\right)^{-\alpha}$ approximately describes the linewidth with $\alpha = 1$ in the 1D case. For temperatures above ~ 50 K (~ 75 K for line II) the ESR linewidth enters to the paramagnetic regime, where the spin-lattice relaxation processes become dominant and a linear low can be used to describe the temperature evolution of linewidth. Thus we have tried to fit $\Delta H(T)$ in the temperature interval 15-300 K by using both mechanisms mentioned above and the equation

$$\Delta H(T) = \Delta H_0 + A \cdot T + B \cdot (T - T_{cr})^{-1},$$

where $\Delta H_0$, $A$, $B$ and $T_{cr}$ are four fitting parameters. The results as the solid lines are shown in the middle panels of Fig. 6. Note, in contrast to the weak resonance, the obtained parameter of $\Delta H_0$ for the broad line is close to zero. The obtained values of $T_{cr}$ well correlates with the value of $T_N$ too.

With decreasing temperature the small shifts of the resonance field $H_{res}$ for both lines are detected (see the low panels in Fig. 6). Due to the low confidence to the observed temperature-dependent shifts any model suggestions or interpretations are not regarded.

## 5. Conclusions

In summary, we have performed a first study of the angular and temperature dependences of the X-band ESR spectra on single crystals of the quasi-1D spin-½ zigzag chain system $\beta$-$TeVO_4$. It was detected two contributions to the ESR signal: an extremely broad resonance around $H_{res} \approx 6.4$ kOe ($\langle g \rangle \approx 1.2$) (the best condition for its registration is $H \perp b$) and a small but narrow resonance on top of the broad signal. Due to the large linewidth and relatively small intensity of the main resonance, we could not observe the typical ESR behavior for conventional 1D Heisenberg antiferromagnet. The resonance field of the dominating broad signal is shifted to the $g$-values ($g_\perp = 1.6$; $g_\parallel = 0.8$) with the average $\langle g \rangle \approx 1.2$ well below $g = 2$ and becomes undetectable upon the onset of strong 1D correlations below 15 K. The second signal, which is observed for all orientations with a rather small, nearly invariable, linewidth and intensity, is in agreement with a very weak anisotropy of an $S$-state ion, that is expected for the $V^{4+}$ ions in a crystal field environment. But the weak resonance might be due to an impurity. Therefore, compound $\beta$-$TeVO_4$ is a exceptional example of a material with the $V^{4+}$ ions ($S = ½$) in a square-pyramidal oxygen environment for which a resonance signal is difficult to detect and interpret.

In the future, more detailed analysis will be performed to estimate of the nature of the individual mechanisms responsible for magnetic anisotropy and the linewidth broadening.